\documentstyle[preprint,prd,aps,epsf,axodraw]{revtex}
\tightenlines
\newcommand{\NP}{Nucl. Phys. }
\newcommand{\PR}{Phys. Rev. }
\newcommand{\PRL}{Phys. Rev. Lett. }
\newcommand{\PL}{Phys. Lett. }

\begin{document}

\pagenumbering{arabic}

\begin{flushright}
TUIMP-TH-99/104

AS-ITP-99-08
\end{flushright}

\begin{center}
{\Large\sf Vanishing Contribution to quark electric dipole moment
in the 2HD model with CKM CP violation}
\\[10pt]
\vspace{1.0 cm}

{Yi Liao}
\vspace{1.5ex}

{\small Department of Modern Applied Physics, Tsinghua University,
Beijing 100084, P.R.China\\}
\vspace{3.0ex}
{Xiaoyuan Li}
\vspace{1.5ex}

{\small Institute of Theoretical Physics, The Chinese Academy of Sciences,
Beijing 100080, P.R.China\\}
\vspace{3.0ex}
{\bf Abstract}
\end{center}

In the standard model ( SM ) of electroweak interactions, CP noninvariance
arises from the nonzero phase in the CKM matrix. Its contribution to the
quark electric dipole moment ( EDM ) vanishes surprisingly at two
loop order. This makes the quark EDM extremely small in the SM. In this
paper, we consider the two Higgs doublet extension of the SM and assume
that CP noninvariance is still encoded in the CKM matrix. We calculate
the charged Higgs boson contribution to the quark EDM which naively should
be of order $eG_F^2\tilde{\delta}(4\pi)^{-4}m_{u(d)}m_t^2m_b^2m_H^{-2}$ for
the up ( down ) quark with possible enhancement factors of $\tan^2\beta$.
Here $\tilde{\delta}$ is the rephasing invariant of CP violation.
However, contrary to the above naive expectation, we find that the charged
Higgs boson contribution vanishes strictly at two loop order. We show
explicitly how this comes about and explains how it is related to the
general form of Yukawa couplings in a spontaneously broken gauge theory.

\begin{flushleft}
\end{flushleft}

\begin{flushleft}
\end{flushleft}

\vspace{1cm}
\begin{center}
             submitted to {\it Phys. Rev. D}
\end{center}

\newpage

\section{Introduction}

CP noninvariance in any models with CPT symmetry will in general induce
P- and T-violating electric dipole moments ( EDM ) for elementary particles
through quantum effects. The discovery of these moments would be direct
evidence of CP noninvariance outside the scope of the neutral kaon system
and would help us identify the origin of CP noninvariance$\cite{review}$.
Of special interest among these moments is the neutron EDM.
The current experimental upper bound for the neutron is
$|d(N)|<1.1\times 10^{-25} {\rm ~e~cm}$ $\cite{neutron}$
and this limit is hopefully to be improved by several orders of magnitude
in the near future$\cite{future}$.

From theoretical point of view, any calculation of the neutron EDM proceeds
in two steps. In the first step, one writes down all relevant operators
which break the CP symmetry and involve only light degrees of freedom.
These operators are usually defined at a high energy scale where CP
violation occurs. They include the electric or chromo-electric dipole
moments of the light quarks and gluons$\cite{weinberg}$
, four quark operators$\cite{kky}$ and possibly
others. In the second step, they are evolved down to the typical hadronic
scale and their effects on the neutron EDM are then calculated. While
naive dimensional analysis or hadronic models have to be invoked in the
second step, the first step can be implemented unambiguously once
the model of CP violation is specified. In this work, we will be concerned
with this first step calculation, in particular, the calculation of the
quark EDM.

In the standard model ( SM ) of electroweak interactions, CP noninvariance
arises from the nonzero phase in the CKM matrix. The quark EDM vanishes
trivially at one loop order since only moduli of the matrix are involved
in the relevant amplitude. At two loop order, the flavour structure is
rich enough which could in principle allow for a CP-violating
EDM$\cite{ellis}$; however,
the final contribution to the quark EDM vanishes surprisingly when the
sum over internal virtual flavours is taken$\cite{shab}\cite{donoghue}$.
This circumstance also appears in the $W^{\pm}$ EDM
$\cite{kp91}\cite{booth}$. Although there are attempts to understand the
vanishing result$\cite{khrip}\cite{ck}$, it seems clear by now that
it does not result from any symmetry which would dictate the zero EDM
automatically at the lowest nontrivial order. Furthermore the vanishing
result is accidental in some sense, as a result of specific Lorentz
and flavour structure. First, the $W^{\pm}$ boson still acquires a
nonvanishing, P- and T-violating magnetic quadrupole moment at two loop
order$\cite{kp94}$ though its EDM vanishes at the same
order. Second, the quark EDM does not vanish any more when QCD corrections
are included$\cite{shab80}\cite{khrip}\cite{ck97}$.
Of course, this makes the quark EDM extremely small in the
SM. If the light quark EDM is one of the important contributions to the
neutron EDM, it is then hopeless to observe the neutron EDM in the near
future. Considering this, we would like to investigate how the quark
EDM could be enhanced beyond the SM. In this paper we study the two Higgs
doublet extension of the SM$\cite{higgs}$. We will assume conservatively
that CP noninvariance is still encoded in the CKM matrix so that the
charged Higgs boson $H^{\pm}$ is the only other particle besides
$W^{\pm}$ that mediates CP violation. Since the Yukawa couplings
between $H^{\pm}$ and quarks generally involve the relevant quark
masses and are thus less universal as compared to the gauge couplings
between $W^{\pm}$ and quarks, we would expect naively that the
$H^{\pm}$ exchange will make a contribution to the quark EDM which
should be of order
$d(u(d))\sim eG_F^2\tilde{\delta}(4\pi)^{-4}m_{u(d)}m_t^2m_b^2m_H^{-2}$
for the up ( down ) quark. [ Here $\tilde{\delta}$ is the rephasing
invariant of CP violation$\cite{jarlskog}$. ]
As we will display later on, $d(d)$
could even be enhanced by a large factor of $\tan^2\beta$ which
would make $d(d)$ easily reach the level of $10^{-31}{\rm ~e~cm}$
for a charged Higgs mass
of 200 GeV. However, a detailed calculation shows that the above naive
expectation is actually not realized in the final result: the $H^{\pm}$
contribution vanishes strictly at two loop order when the sum over
internal virtual quark flavours is taken. We show explicitly how this null
result comes about as a consequence of the general form of Yukawa couplings
in a spontaneously broken gauge theory.

The following sections are organized as follows. In section 2, we first
discuss the renormalization of one loop elements to be used in the
complete two loop calculation. Then, we present respectively the
contributions from exchanges of two charged Higgs bosons, one charged
Higgs boson and one $W^{\pm}$, for the general form of Yukawa couplings.
In passing we also give the result from exchanges of two $W^{\pm}$ which
was previously calculated in the SM. The naive expectation for $d(u)$ and
$d(d)$ is then verified. In section 3, we first show that the
contribution involving exchanges of $H^{\pm}$ or $W^{\pm}$ vanishes
when we sum over internal quark flavours.
We also indicate a difference of the cancellation mechanism in the
present case and in the SM case as computed in the unitarity gauge.
Then
we examine generally how it could be possible to have such
a vanishing result. Section 4 is a recapitulation of our result.

\section{Explicit result of charged Higgs boson contributions}

We shall calculate in this section the quark EDM arising from exchange of
$H^{\pm}$ and $W^{\pm}$ in the two Higgs doublet extension of the SM. The
Feynman diagrams at the lowest two loop order are depicted in Fig. 1. We
shall denote the internal up-type quarks by Greek letters $\alpha,~\beta$
etc, the internal down-type quarks by Latin letters $j,~k$ etc, and the
external up- or down-type quark by $e$. Within this section, the flavours
of internal quarks are fixed. We shall examine in the next section what
will happen when summation over flavours is done.

To set up our notation, we first list the relevant Feynman rules. The
Feynman rule for the $\bar{u}_{\alpha}d_j W_{\mu}^+$ vertex is
\begin{equation}
\displaystyle i\frac{g}{2\sqrt{2}}V_{\alpha j}\gamma_{\mu}(1-\gamma_5),
\end{equation}
where $g$ is the $SU(2)$ weak coupling constant and $V_{\alpha j}$ is the
entry $(\alpha,j)$ of the CKM matrix. Then the vertex $\bar{d}_ju_{\alpha}
W_{\mu}^-$ is
$\displaystyle i\frac{g}{2\sqrt{2}}V_{\alpha j}^{\star}
\gamma_{\mu}(1-\gamma_5)$.
The Feynman rule for the $\bar{u}_{\alpha}d_j H^+$ vertex is parametrized
as
\begin{equation}
\displaystyle i\frac{g}{2\sqrt{2}m_W}V_{\alpha j}
(C_{\alpha j}+C_{\alpha j}^{\prime}\gamma_5),
\end{equation}
where $C_{\alpha j}$ and $C_{\alpha j}^{\prime}$ are real constants and
may depend on the masses of $u_{\alpha}$ and $d_j$. The vertex
$\bar{d}_ju_{\alpha}H^-$ is then
$\displaystyle i\frac{g}{2\sqrt{2}m_W}V_{\alpha j}^{\star}
(C_{\alpha j}-C_{\alpha j}^{\prime}\gamma_5)$.
We emphasize again that $C_{\alpha j}$ and $C_{\alpha j}^{\prime}$ are
assumed to be real in our calculation; i.e., CP noninvariance occurs only
in the CKM matrix. If they are complex numbers, it will be completely
another story$\cite{ww}$. The ordinary couplings in the Minimal Supersymmetric
Standard Model ( MSSM ) are recovered by setting
\begin{equation}
\displaystyle C_{\alpha j}=u_{\alpha}\cot\beta+d_j\tan\beta,
~C_{\alpha j}^{\prime}=-u_{\alpha}\cot\beta+d_j\tan\beta,
\end{equation}
where $\tan\beta$ is a parameter measuring the ratio of the vacuum
expectation values of the two Higgs doublet fields. From now on, we
always use the names of quarks to denote their masses. The vertices
involving the would-be Goldstone bosons, $\bar{u}_{\alpha}d_j G^+$
and $\bar{d}_ju_{\alpha}G^-$, also arise as a special case:
\begin{equation}
\displaystyle C_{\alpha j}=u_{\alpha}-d_j,
~C_{\alpha j}^{\prime}=-u_{\alpha}-d_j.
\end{equation}
To simplify the computation of diagrams involving $W^{\pm}$ exchange,
we shall use the background field gauge
$\cite{abbott}\cite{einhorn}\cite{ll}$
( or the nonlinear $R_{\xi}$ gauge$\cite{nonlinear}$ )
with $\xi=1$. There will be no mixed $W^{\pm}G^{\mp}A$ vertex
[ $A$ is the external electromagnetic field ], and the Feynman rule for
the $W^+_{\rho}W^-_{\sigma}A_{\mu}$ vertex is
\begin{equation}
\displaystyle -ie[(k_0-k_+)_{\sigma}g_{\mu\rho}+(k_+-k_-)_{\mu}
g_{\rho\sigma}+(k_--k_0)_{\rho}g_{\sigma\mu}],
\end{equation}
where $k_0,~k_+,~k_-$ are incoming momenta for the fields $A_{\mu},
~W_{\rho}^+,~W_{\sigma}^-$. Finally, to avoid any ambiguity, we define
the effective EDM interaction as
\begin{equation}
\displaystyle{\cal L}_{\rm eff}=id\bar{\psi}\gamma_5\sigma_{\mu\nu}\psi
F^{\mu\nu},
\end{equation}
where $d$ is the EDM of the fermion $\psi$ and it is real by Hermiticity.
The Feynman rule for the effective vertex $\bar{\psi}\psi A_{\mu}$ is
\begin{equation}
\displaystyle -d[\gamma_{\mu},\rlap/q]\gamma_5,
\end{equation}
where $q$ is the outgoing momentum of the photon.

\noindent{\large A. Renormalization of one loop elements}

We shall be mainly concerned with the EDM of the up-type quark. The result
for the down-type quark will be obtained by simple substitutions at the
end. To induce an EDM for the quark $u_e$, the imaginary part of the CKM
matrix must be involved so that the flavours $j$ and $k$ in Fig. 1 are
different. Therefore, we need to renormalize the off-diagonal self-energy
$-i\Sigma^{kj}(\ell)$ and the vertex with the photon
$ie\Gamma_{\mu}^{kj}(\ell,\ell+q;q)$. Denote the bare one loop contribution
by a hat and the counter-term one by a tilde, so that the
renormalized quantities are
\begin{equation}
\displaystyle\Sigma^{kj}=\hat{\Sigma}^{kj}+\tilde{\Sigma}^{kj},~
\Gamma_{\mu}^{kj}=\hat{\Gamma}_{\mu}^{kj}+\tilde{\Gamma}_{\mu}^{kj}.
\end{equation}
The general structure of $\hat{\Sigma}^{kj}$ may be parametrized as
\begin{equation}
\displaystyle\hat{\Sigma}^{kj}(\ell)=\rlap/\ell[A(\ell^2)+B(\ell^2)
\gamma_5]+u_{\alpha}[C(\ell^2)+D(\ell^2)\gamma_5].
\end{equation}
The off-diagonal self-energy is renormalized by requiring that there be
no mixing when either of $d_j$ and $d_k$ is on-shell:
\begin{equation}
\displaystyle\bar{d}_k(\ell)\Sigma^{kj}(\ell)|_{\rlap/\ell=d_k}=0,~
\Sigma^{kj}(\ell)d_j(\ell)|_{\rlap/\ell=d_j}=0.
\end{equation}
The counter-term is then determined to be
\begin{equation}
\begin{array}{l}
\displaystyle\tilde{\Sigma}^{kj}(\ell)=\frac{1}{d_k-d_j}
\left[(\rlap/\ell-d_k)\left(d_jA(d_j^2)+u_{\alpha}C(d_j^2)\right)-
\left(d_kA(d_k^2)+u_{\alpha}C(d_k^2)\right)(\rlap/\ell-d_j)\right]\\
\displaystyle +\frac{1}{d_k+d_j}
\left[(\rlap/\ell-d_k)\gamma_5\left(-d_jB(d_j^2)+u_{\alpha}D(d_j^2)
\right)
+\left(d_kB(d_k^2)+u_{\alpha}D(d_k^2)\right)\gamma_5
(\rlap/\ell-d_j)\right].
\end{array}
\end{equation}
The renormalization of the vertex $\Gamma_{\mu}^{kj}$ is not independent
but related to that of the self-energy by the Ward identity,
\begin{equation}
\displaystyle q^{\mu}ie\Gamma_{\mu}^{kj}(\ell,\ell+q;q)=
ieQ_d\left[\Sigma^{kj}(\ell)-\Sigma^{kj}(\ell+q)\right].
\end{equation}
It may be explicitly checked that the bare one loop quantities satisfy
the above identity so that the latter must also be separately satisfied
by the counter-term quantities. In this way, we find
\begin{equation}
\begin{array}{l}
\displaystyle\tilde{\Gamma}_{\mu}^{kj}(\ell,\ell+q;q)=
Q_d\gamma_{\mu}\left\{\frac{1}{d_k-d_j}
\left[\left(d_kA(d_k^2)-d_jA(d_j^2)\right)
+u_{\alpha}\left(C(d_k^2)-C(d_j^2)\right)\right]\right.\\
\displaystyle +\left.\frac{1}{d_k+d_j}
\left[\left(d_kB(d_k^2)+d_jB(d_j^2)\right)
+u_{\alpha}\left(D(d_k^2)-D(d_j^2)\right)\right]\gamma_5\right\}.
\end{array}
\end{equation}
Since we shall present separate contributions from exchanges of $H^{\pm}$
and $W^{\pm}$ in the subsequent subsections, we give below the functions
$A,~B,~C,~D$ arising from exchanges of $H^{\pm},~W^{\pm}$ and $G^{\pm}$.
We work in $n=4-2\epsilon$ dimensions to regularize ultraviolet
divergences. For the $H^{\pm}$ exchange,
\begin{equation}
\begin{array}{rcl}
\displaystyle A(\ell^2)&=&
-(4\pi)^{-2}G_F/\sqrt{2}V_{\alpha k}^{\star}V_{\alpha j}
(C_{\alpha k}C_{\alpha j}+C_{\alpha k}^{\prime}C_{\alpha j}^{\prime})
f_1(\ell^2,u_{\alpha}^2,m_H^2),\\
\displaystyle B(\ell^2)&=&
-(4\pi)^{-2}G_F/\sqrt{2}V_{\alpha k}^{\star}V_{\alpha j}
(C_{\alpha k}C_{\alpha j}^{\prime}+C_{\alpha k}^{\prime}C_{\alpha j})
f_1(\ell^2,u_{\alpha}^2,m_H^2),\\
\displaystyle C(\ell^2)&=&
-(4\pi)^{-2}G_F/\sqrt{2}V_{\alpha k}^{\star}V_{\alpha j}
(C_{\alpha k}C_{\alpha j}-C_{\alpha k}^{\prime}C_{\alpha j}^{\prime})
f_0(\ell^2,u_{\alpha}^2,m_H^2),\\
\displaystyle D(\ell^2)&=&
-(4\pi)^{-2}G_F/\sqrt{2}V_{\alpha k}^{\star}V_{\alpha j}
(C_{\alpha k}C_{\alpha j}^{\prime}-C_{\alpha k}^{\prime}C_{\alpha j})
f_0(\ell^2,u_{\alpha}^2,m_H^2).
\end{array}
\end{equation}
For the $W^{\pm}$ exchange in the $\xi=1$ gauge,
\begin{equation}
\begin{array}{l}
\displaystyle A(\ell^2)=-B(\ell^2)=+(4\pi)^{-2}g^2
V_{\alpha k}^{\star}V_{\alpha j}\frac{1}{4}(2-n)
f_1(\ell^2,u_{\alpha}^2,m_W^2),\\
C(\ell^2)=D(\ell^2)=0.
\end{array}
\end{equation}
The $G^{\pm}$ exchange in $\xi=1$ gauge is a special case of the
$H^{\pm}$ exchange, i.e., $m_H^2\to m_W^2$, and with couplings $C$ and
$C^{\prime}$ substituted by values in Eqn.(4). The functions $f_1$ and
$f_0$ are given in the Appendix.

\noindent{\large B. Double $H^{\pm}$ exchanges}

The momentum arrangement for external quarks and photon is shown in Fig.1.
To pick out the EDM, we expand to the linear order term in the photon
momentum $q$. One should be careful in dropping terms that are
superficially of zero order in $q$, since some of them are actually
of linear order when the equation of motion is applied, and thus may
contribute to the EDM. Notice that the final $\gamma_5$ in the effective
EDM vertex can only come from Yukawa vertices since there would be no
P violation if no $\gamma_5$ were involved in these vertices. To
simplify the expression, the equation of motion for external quarks
is freely used and only those terms that can finally contribute are
kept. After a tedious computation, the Feynman diagrams in Fig. 1 sum
up to the following structure with a common coefficient,
$(4\pi)^{-2}eG_F^2/2V_{ek}V_{\alpha k}^{\star}V_{\alpha j}V_{ej}^{\star}
q^{\nu}\gamma_5$:
\begin{equation}
\begin{array}{l}
\displaystyle
+C_1(e\alpha;kj)\left[F_{1,\mu\nu}(k)-F_{1,\mu\nu}(j)\right]
+C_2(e\alpha;kj)\left[F_{2,\mu\nu}(k)-F_{2,\mu\nu}(j)\right]\\
\displaystyle
+C_3(e\alpha;kj)F_{3,\mu\nu}(k)-C_3(e\alpha;jk)F_{3,\mu\nu}(j)\\
\displaystyle
+C_4(e\alpha;kj)\left[F_{4,\mu\nu}(k)-F_{4,\mu\nu}(j)\right]
+C_5(e\alpha;kj)\left[F_{5,\mu\nu}(k)-F_{5,\mu\nu}(j)\right],
\end{array}
\end{equation}
where $C_i$ are combinations of Yukawa couplings and quark masses $d_k$
and $d_j$:
\begin{equation}
\begin{array}{rcl}
\displaystyle C_1(e\alpha;kj)&=&\left[
(C_{ek}C_{ej}-C_{ek}^{\prime}C_{ej}^{\prime})
(C_{\alpha k}^{\prime}C_{\alpha j}-C_{\alpha k}C_{\alpha j}^{\prime})
\right.\\
&&\displaystyle
-\left.(C_{\alpha k}C_{\alpha j}-C_{\alpha k}^{\prime}C_{\alpha j}^{\prime})
(C_{ek}^{\prime}C_{ej}-C_{ek}C_{ej}^{\prime})\right]
\frac{d_jd_k}{d_k^2-d_j^2},\\
\displaystyle C_2(e\alpha;kj)&=&\left[
(C_{ek}C_{ej}-C_{ek}^{\prime}C_{ej}^{\prime})
(C_{\alpha k}^{\prime}C_{\alpha j}-C_{\alpha k}C_{\alpha j}^{\prime})
\right.\\
&&\displaystyle+\left.
(C_{\alpha k}C_{\alpha j}-C_{\alpha k}^{\prime}C_{\alpha j}^{\prime})
(C_{ek}^{\prime}C_{ej}-C_{ek}C_{ej}^{\prime})
\right]\frac{1}{d_k^2-d_j^2},\\
\displaystyle C_3(e\alpha;kj)&=&\displaystyle \left[-
(C_{ek}C_{ej}-C_{ek}^{\prime}C_{ej}^{\prime})
(C_{\alpha k}^{\prime}C_{\alpha j}-C_{\alpha k}C_{\alpha j}^{\prime})
\frac{1}{d_k+d_j}\right.\\
&&\displaystyle-\left.
(C_{\alpha k}C_{\alpha j}-C_{\alpha k}^{\prime}C_{\alpha j}^{\prime})
(C_{ek}^{\prime}C_{ej}-C_{ek}C_{ej}^{\prime})\frac{1}{d_k-d_j}
\right]d_k,\\
\displaystyle C_3(e\alpha;jk)&=&C_3(e\alpha;kj)|_{j\leftrightarrow k},\\
\displaystyle C_4(e\alpha;kj)&=&\displaystyle \left[-
(C_{ek}C_{ej}-C_{ek}^{\prime}C_{ej}^{\prime})
(C_{\alpha k}^{\prime}C_{\alpha j}+C_{\alpha k}C_{\alpha j}^{\prime})
\frac{1}{d_k+d_j}\right.\\
&&\displaystyle+\left.
(C_{ek}^{\prime}C_{ej}-C_{ek}C_{ej}^{\prime})
(C_{\alpha k}C_{\alpha j}+C_{\alpha k}^{\prime}C_{\alpha j}^{\prime})
\frac{1}{d_k-d_j}\right],\\
\displaystyle C_5(e\alpha;kj)&=&C_4(e\alpha;kj)|_{\alpha\leftrightarrow e}.
\end{array}
\end{equation}
The functions $F_{i,\mu\nu}$ are complicated loop momentum integrals
which are too lengthy to be displayed here. For
brevity, we only indicate their dependence on the internal quark mass
$d_k$ or $d_j$ although they depend as well on the external momentum
$p$, the internal quark mass $u_{\alpha}$ and the charged Higgs
mass $m_H$.

The above results are obtained without using any approximations.
The discussions in the next section will be based on these general
results. We notice from Eqn.(17) that the coupling combinations
$C_1,~C_2,~C_4$ and $C_5$ are symmetric with respect to $k$ and $j$
such that the structure in Eqn.(16) is antisymmetric with respect to
$k$ and $j$. Therefore, the mirror-reflected diagrams corresponding
to interchange of $k$ and $j$ in Fig.1 are simply related by
$\displaystyle V_{ek}V_{\alpha k}^{\star}V_{\alpha j}V_{ej}^{\star}\to
-V_{ej}V_{\alpha j}^{\star}V_{\alpha k}V_{ek}^{\star}
=-\left(V_{ek}V_{\alpha k}^{\star}V_{\alpha j}V_{ej}^{\star}\right)
^{\star}$,
so that in their sum the
${\rm Re}(V_{ek}V_{\alpha k}^{\star}V_{\alpha j}V_{ej}^{\star})$ term
is cancelled while the
${\rm Im}(V_{ek}V_{\alpha k}^{\star}V_{\alpha j}V_{ej}^{\star})$ term
is doubled. This is essential, as emphasized for the SM case
in Ref.$\cite{ck}$,
to guarantee that the EDM is a real number as required by the
Hermiticity of the effective action. The results for the down-type
quark $d_e$ are obtained by the following substitutions:
\begin{equation}
\begin{array}{l}
\displaystyle V_{ek}V_{\alpha k}^{\star}V_{\alpha j}V_{ej}^{\star}\to
V_{\beta e}^{\star}V_{\beta i}V_{\alpha i}^{\star}V_{\alpha e},\\
\displaystyle Q_u\leftrightarrow Q_d,\\
\displaystyle C_{ek}\to C_{\beta e},~C_{\alpha k}\to C_{\beta i},~
C_{\alpha j}\to C_{\alpha i},~C_{ej}\to C_{\alpha e},\\
\displaystyle C_{ek}^{\prime}\to -C_{\beta e}^{\prime},
~C_{\alpha k}^{\prime}\to -C_{\beta i}^{\prime},
~C_{\alpha j}^{\prime}\to -C_{\alpha i}^{\prime},
~C_{ej}^{\prime}\to -C_{\alpha e}^{\prime},\\
u_{\alpha}\to d_i,~d_j\to u_{\alpha},~d_k\to u_{\beta},~u_e\to d_e.
\end{array}
\end{equation}

Before concluding this subsection, we would like to get some idea of
how the quark EDM looks like. For this purpose, let us specialize to
the case of the $u$ and $d$ quarks in the MSSM. We may use then the
small external mass approximation ( SEMA ). In this approximation,
only terms linear in the external mass are kept while higher order
terms are safely ignored. [ At least one factor of external mass is
involved due to the chirality flip feature of the EDM operator. ]
The formula simplifies considerably. Fig. 1 along with its mirror
reflection gives for the $u$ quark,
\begin{equation}
\begin{array}{rcl}
\displaystyle d(u)&=&+eG_F^2{\rm Im}
(V_{uk}V_{\alpha k}^{\star}V_{\alpha j}V_{uj}^{\star})
(4\pi)^{-2}4uu_{\alpha}^2[F(k)-F(j)],\\
F(k)&=&\displaystyle +Q_u i\int\frac{d^4\ell}{(2\pi)^4}\left[
\frac{m_H^2d_k^2}{D_H^3D_k}\left(I_1(k)\cot^2\beta+I_0(k)\right)+
\frac{\ell^2}{D_HD_k}\frac{1}{2}J_{1,1}\cot^2\beta\right]\\
&&\displaystyle +Q_d i\int\frac{d^4\ell}{(2\pi)^4}\left[
\frac{d_k^2}{D_HD_k^2}\left(I_1(k)\cot^2\beta+I_0(k)\right)+
\frac{\ell^2}{D_HD_k}\left(\frac{1}{2}J_{2,1}\cot^2\beta+J_{1,1}\right)
\right],
\end{array}
\end{equation}
where $D_H=\ell^2-m_H^2$. For later discussion we should mention that
the whole contribution from Fig. 1(a) is given by the term involving
$D_H^{-3}$ upon setting $Q_u$ to 1.
The $I-$ and $J-$functions arise from the inner loop integration
and their explicit forms are given in the Appendix. The relevant feature
at the moment is that the $J-$functions depend only on $\ell^2$ and the
inner loop masses $m_H^2$ and $u_{\alpha}^2$ while the $I-$functions
depend also on $d_k^2$ or $d_j^2$ as indicated above.
For the $d$ quark,
\begin{equation}
\begin{array}{rcl}
\displaystyle d(d)&=&+eG_F^2{\rm Im}
(V_{\beta d}^{\star}V_{\beta i}V_{\alpha i}^{\star}V_{\alpha d})
(4\pi)^{-2}4dd_i^2[F(\beta)-F(\alpha)],\\
F(\beta)&=&\displaystyle +Q_d i\int\frac{d^4\ell}{(2\pi)^4}\left[
\frac{m_H^2u_{\beta}^2}{D_H^3D_{\beta}}
\left(I_1(\beta)\tan^2\beta+I_0(\beta)\right)+
\frac{\ell^2}{D_HD_{\beta}}\frac{1}{2}J_{1,1}\tan^2\beta\right]\\
&&\displaystyle +Q_u i\int\frac{d^4\ell}{(2\pi)^4}\left[
\frac{u_{\beta}^2}{D_HD_{\beta}^2}
\left(I_1(\beta)\tan^2\beta+I_0(\beta)\right)+
\frac{\ell^2}{D_HD_{\beta}}\left(\frac{1}{2}J_{2,1}\tan^2\beta+J_{1,1}\right)
\right],
\end{array}
\end{equation}
where $D_{\beta}=\ell^2-u_{\beta}^2$, $J_{1,1}$ and $J_{2,1}$ depend on
$\ell^2,~d_i^2$ and $m_H^2$ while $I_0(\beta)$ and $I_1(\beta)$ depend
on $u_{\beta}^2$ as well. Suppose that $u_{\alpha}$, $d_k=d_i$ and $d_j$
are respectively the top, bottom and down quarks, we find that, up to
logarithms,
\begin{equation}
\begin{array}{rcl}
d(u)&\sim&\displaystyle  eG_F^2\tilde{\delta}(4\pi)^{-4}m_um_t^2
\frac{m_b^2-m_d^2}{m_H^2}\cdot\left(1{\rm ~and~}\cot^2\beta\right),\\
d(d)&\sim&\displaystyle  eG_F^2\tilde{\delta}(4\pi)^{-4}m_dm_b^2
\frac{m_t^2}{m_H^2}\cdot\left(1{\rm ~and~}\tan^2\beta\right),
\end{array}
\end{equation}
where
$\tilde{\delta}={\rm Im}(V_{ub}V_{tb}^{\star}V_{td}V_{ud}^{\star})$.
For numerical estimate we take the following input parameters,
$G_F\sim 10^{-5}{\rm ~GeV}^{-2},~|\tilde{\delta}|\sim 5\cdot 10^{-5}$,
$m_u\sim m_d\sim 5{\rm ~MeV},~m_t\sim 170{\rm ~GeV},~m_b\sim 4.5
{\rm ~GeV}$, $m_H\sim 200{\rm ~GeV}$ and $\tan\beta\sim 30$, then,
\begin{equation}
|d(u)|\sim 10^{-34}{\rm ~e~cm},~|d(d)|\sim 10^{-31}{\rm ~e~cm}.
\end{equation}
Fo comparison, we quote the three loop result in the SM$\cite{ck97}$,
$|d(u)|\sim0.35\cdot 10^{-34}{\rm ~e~cm}$,
$|d(d)|\sim0.15\cdot 10^{-34}{\rm ~e~cm}$. Therefore, if not for the
cancellation mechanism to be discussed in the next section, the result
for the light quark EDM in the two Higgs doublet model would be very
different from that in the SM.

\noindent{\large C. Outer loop $W^{\pm}$ plus inner loop $H^{\pm}$ exchanges}

Since we work in the background field gauge, we present the separate
results from $W^{\pm}$ and $G^{\pm}$ exchanges. For the $W^{\pm}$
exchange, Fig. 1(a) does not contribute to the EDM and Fig. 1(b) is
completely cancelled by corresponding terms in diagrams (c)-(e).
The counter-term diagrams do not contribute either. Diagrams (c)-(e)
with their mirror reflection then give for the external $u_e$ quark,
\begin{equation}
\begin{array}{l}
\displaystyle+eg^2\sqrt{2}G_F{\rm Im}(V_{ek}V_{\alpha k}^{\star}
V_{\alpha j}V_{ej}^{\star})(4\pi)^{-2}u_{\alpha}\left[
\frac{C_{\alpha k}C_{\alpha j}-C_{\alpha k}^{\prime}C_{\alpha j}^{\prime}}
{d_k+d_j}-
\frac{C_{\alpha k}C_{\alpha j}^{\prime}-C_{\alpha k}^{\prime}C_{\alpha j}}
{d_k-d_j}\right]\\
\displaystyle
\frac{1}{2}q^{\nu}\gamma_5 i\int\frac{d^4\ell}{(2\pi)^4}\frac{1}{D_W}
\left(\frac{1}{D_k}-\frac{1}{D_j}\right)\left(Q_uJ_{0,1}-Q_dJ_{1,1}\right)
(\ell_{\mu}\gamma_{\nu}-\ell_{\nu}\gamma_{\mu}),
\end{array}
\end{equation}
where $D_W=(\ell-p)^2-m_W^2$. The result for the external $d_e$ quark
is obtained by substitutions. In the SEMA, we have
\begin{equation}
\begin{array}{rcl}
d(u_e)&=&\displaystyle +eg^2G_F/\sqrt{2}(4\pi)^{-2}
{\rm Im}(V_{ek}V_{\alpha k}^{\star}V_{\alpha j}V_{ej}^{\star})
\frac{1}{4}u_{\alpha}u_e\\
&&\displaystyle\left[
\frac{C_{\alpha k}C_{\alpha j}-C_{\alpha k}^{\prime}C_{\alpha j}^{\prime}}
{d_k+d_j}-
\frac{C_{\alpha k}C_{\alpha j}^{\prime}-C_{\alpha k}^{\prime}C_{\alpha j}}
{d_k-d_j}\right]\\
&&\displaystyle i\int\frac{d^4\ell}{(2\pi)^4}\frac{\ell^2}{D_W^2}
\left(\frac{1}{D_k}-\frac{1}{D_j}\right)\left(Q_uJ_{0,1}-Q_dJ_{1,1}
\right),
\end{array}
\end{equation}
where now $D_W=\ell^2-m_W^2$. The $G^{\pm}$ contribution is a special
case of the double $H^{\pm}$ exchange; i.e., we only need to replace
the couplings $C_{ek},~C_{ej},~C_{ek}^{\prime},~C_{ej}^{\prime}$ by
their values in Eqn.(4) and $D_H$ by $D_W$. We display here the
coupling combinations which are relevant to discussions in the next
section:
\begin{equation}
\begin{array}{rcl}
C_1(e\alpha;kj)&=&\displaystyle 2u_ed_kd_j\left[
\frac{C_{\alpha k}C_{\alpha j}-
C_{\alpha k}^{\prime}C_{\alpha j}^{\prime}}{d_k+d_j}+
\frac{C_{\alpha k}C_{\alpha j}^{\prime}-
C_{\alpha k}^{\prime}C_{\alpha j}}{d_k-d_j}\right],\\
C_2(e\alpha;kj)&=&\displaystyle -2u_e\left[
\frac{C_{\alpha k}C_{\alpha j}-
C_{\alpha k}^{\prime}C_{\alpha j}^{\prime}}{d_k+d_j}-
\frac{C_{\alpha k}C_{\alpha j}^{\prime}-
C_{\alpha k}^{\prime}C_{\alpha j}}{d_k-d_j}\right],\\
C_3(e\alpha;kj)&=&\displaystyle 2u_ed_k[
(C_{\alpha k}C_{\alpha j}-
C_{\alpha k}^{\prime}C_{\alpha j}^{\prime})-
(C_{\alpha k}C_{\alpha j}^{\prime}-
C_{\alpha k}^{\prime}C_{\alpha j})],\\
C_4(e\alpha;kj)&=&\displaystyle -2u_e
(C_{\alpha k}-C_{\alpha k}^{\prime})
(C_{\alpha j}-C_{\alpha j}^{\prime}),\\
C_5(e\alpha;kj)&=&\displaystyle 2\left[
(C_{\alpha k}C_{\alpha j}-C_{\alpha k}^{\prime}C_{\alpha j}^{\prime}
)\frac{u_e^2-d_kd_j}{d_k+d_j}+
(C_{\alpha k}^{\prime}C_{\alpha j}-C_{\alpha k}C_{\alpha j}^{\prime}
)\frac{u_e^2+d_kd_j}{d_k-d_j}\right].
\end{array}
\end{equation}

\noindent{\large D. Outer loop $H^{\pm}$ plus inner loop $W^{\pm}$ exchanges}

The contribution from $W^{\pm}$ exchange in Fig. 1 and its mirror
reflection is
\begin{equation}
\begin{array}{l}
\displaystyle+eg^2\sqrt{2}G_F{\rm Im}(V_{ek}V_{\alpha k}^{\star}
V_{\alpha j}V_{ej}^{\star})(4\pi)^{-2}
\left[\frac{C_{ek}C_{ej}-C_{ek}^{\prime}C_{ej}^{\prime}}{d_k+d_j}-
\frac{C_{ek}C_{ej}^{\prime}-C_{ek}^{\prime}C_{ej}}
{d_k-d_j}\right]\\
\displaystyle
\frac{1}{4}q^{\nu}\gamma_5 \left[H_{\mu\nu}(k)-H_{\mu\nu}(j)\right],
\end{array}
\end{equation}
where $H_{\mu\nu}$ is another chain of loop momentum integrals.
In the SEMA, we have for the $u_e$ quark,
\begin{equation}
\begin{array}{rcl}
\displaystyle d(u_e)&=&+eg^2\sqrt{2}G_F(4\pi)^{-2}
{\rm Im}(V_{ek}V_{\alpha k}^{\star}V_{\alpha j}V_{ej}^{\star})\\
&&\displaystyle\frac{1}{4}\left[
\frac{C_{ek}C_{ej}-C_{ek}^{\prime}C_{ej}^{\prime}}{d_k+d_j}-
\frac{C_{ek}C_{ej}^{\prime}-C_{ek}^{\prime}C_{ej}}
{d_k-d_j}\right]\left[H(k)-H(j)\right],\\
H(k)&=&\displaystyle+Q_u i\int\frac{d^4\ell}{(2\pi)^4}
\frac{m_H^2d_k^2}{D_H^3D_k}I_1(k)
+Q_d i\int\frac{d^4\ell}{(2\pi)^4}
\frac{d_k^2}{D_HD_k^2}I_1(k)\\
&&\displaystyle+\frac{1}{2} i\int\frac{d^4\ell}{(2\pi)^4}
\frac{\ell^2}{D_HD_k}\left[Q_dJ_{2,1}-Q_uJ_{1,1}\right]\\
&&\displaystyle
-(Q_u-Q_d)i\int\frac{d^4\ell}{(2\pi)^4}\frac{\ell^2}{D_HD_k}J_{2,0},
\end{array}
\end{equation}
where $D_H=\ell^2-m_H^2$. The $G^{\pm}$ contribution arises as a
special case of the double $H^{\pm}$ exchange: $m_H^2$ in the $I-$
and $J-$functions is replaced by $m_W^2$ and the coupling combinations
are given by the following ones,
\begin{equation}
\begin{array}{rcl}
C_1(e\alpha;kj)&=&\displaystyle -2u_{\alpha}d_kd_j\left[
\frac{C_{ek}C_{ej}-C_{ek}^{\prime}C_{ej}^{\prime}}{d_k+d_j}+
\frac{C_{ek}C_{ej}^{\prime}-
C_{ek}^{\prime}C_{ej}}{d_k-d_j}\right],\\
C_2(e\alpha;kj)&=&\displaystyle -2u_{\alpha}\left[
\frac{C_{ek}C_{ej}-C_{ek}^{\prime}C_{ej}^{\prime}}{d_k+d_j}-
\frac{C_{ek}C_{ej}^{\prime}-
C_{ek}^{\prime}C_{ej}}{d_k-d_j}\right],\\
C_3(e\alpha;kj)&=&\displaystyle 2u_{\alpha}d_k\left[
(C_{ek}C_{ej}-C_{ek}^{\prime}C_{ej}^{\prime})
\frac{d_k-d_j}{d_k+d_j}-
(C_{ek}C_{ej}^{\prime}-C_{ek}^{\prime}C_{ej})
\frac{d_k+d_j}{d_k-d_j}\right],\\
C_4(e\alpha;kj)&=&\displaystyle 2\left[
(C_{ek}C_{ej}-C_{ek}^{\prime}C_{ej}^{\prime}
)\frac{u_{\alpha}^2-d_kd_j}{d_k+d_j}+
(C_{ek}^{\prime}C_{ej}-C_{ek}C_{ej}^{\prime}
)\frac{u_{\alpha}^2+d_kd_j}{d_k-d_j}\right],\\
C_5(e\alpha;kj)&=&\displaystyle -2u_{\alpha}
(C_{ek}-C_{ek}^{\prime})(C_{ej}-C_{ej}^{\prime}).
\end{array}
\end{equation}

\noindent{\large E. Double $W^{\pm}$ exchanges}

For completeness, we present in this subsection the result from double
$W^{\pm}$ exchanges, i.e., the SM result. Since we work in the
background field gauge, we may separate four kinds of contributions:
double $W^{\pm}$ exchanges, double $G^{\pm}$ exchanges, outer loop
$W^{\pm}$ plus inner loop $G^{\pm}$ exchanges, and
outer loop $G^{\pm}$ plus inner loop $W^{\pm}$ exchanges. For double
$W^{\pm}$ exchanges, Fig. 1(a) does not contribute and the
contributions from (b)-(e) are completely cancelled. The counter-term
diagrams do not contribute either. The case of double $G^{\pm}$
exchanges is recovered from subsection B by $m_H^2\to m_W^2$ and by
evaluating couplings in terms of Eqn.(4); i.e., the coupling combinations
become
\begin{equation}
\begin{array}{l}
\displaystyle C_1(e\alpha;kj)=0,~~
C_2(e\alpha;kj)=+8u_{\alpha}u_e ,~~
C_3(e\alpha;kj)=-8u_{\alpha}u_ed_k^2,\\
C_4(e\alpha;kj)=-8u_{\alpha}^2u_e,~~
C_5(e\alpha;kj)=-8u_{\alpha}u_e^2.
\end{array}
\end{equation}
The contribution from outer loop $W^{\pm}$ plus inner loop $G^{\pm}$
exchanges is obtained from Eqn.(23) by $m_H^2\to m_W^2$ and replacing
the coupling combination in the square parentheses by $-4u_{\alpha}$.
Similarly, for outer loop $G^{\pm}$ plus inner loop $W^{\pm}$
exchanges, we replace the coupling combination of Eqn.(26) by $-4u_e$.

\section{Analysis of cancellation mechanism}

The SM result for the quark EDM was presented in the subsection 2E. For
fixed internal quark flavours $\alpha,~j,~k$ and external quark $u_e$
and upon summing the pair of reflection-related diagrams, it has the
following structure,
\begin{equation}
\displaystyle
d(u_e)={\rm Im}(V_{ek}V_{\alpha k}^{\star}V_{\alpha j}V_{ej}^{\star})
\left[H(k)-H(j)\right],
\end{equation}
where the two terms depend exclusively on flavours $d_k$ and $d_j$
respectively. [ Of course they also depend on the masses of quarks
$u_e,~u_{\alpha}$ and the exchanged bosons. ] When we evaluate the
contribution involving exchange of charged Higgs bosons in the MSSM
by using the couplings in Eqn.(3), we find that the above structure
is also preserved. Summing over the three down-type flavours
$i,~j,~k$ while fixing the up-type flavour $\alpha$, we arrive at
\begin{equation}
\begin{array}{rcl}
\displaystyle d(u_e)&=&+
{\rm Im}(V_{ek}V_{\alpha k}^{\star}V_{\alpha j}V_{ej}^{\star})
\left[H(k)-H(j)\right]+
{\rm Im}(V_{ei}V_{\alpha i}^{\star}V_{\alpha k}V_{ek}^{\star})
\left[H(i)-H(k)\right]\\
\displaystyle &&+
{\rm Im}(V_{ej}V_{\alpha j}^{\star}V_{\alpha i}V_{ei}^{\star})
\left[H(j)-H(i)\right]\\
\displaystyle &=&0,
\end{array}
\end{equation}
where the second equality is due to unitarity of the CKM matrix; e.g.,
the $H(k)$ term is
\begin{equation}
\begin{array}{rl}
\displaystyle &\left[
{\rm Im}(V_{ek}V_{\alpha k}^{\star}V_{\alpha j}V_{ej}^{\star})-
{\rm Im}(V_{ei}V_{\alpha i}^{\star}V_{\alpha k}V_{ek}^{\star})
\right]H(k)\\
=&{\rm Im}\left[V_{ek}V_{\alpha k}^{\star}
(\delta_{\alpha e}-V_{\alpha k}V_{ek}^{\star})\right]~H(k)\\
=&0.
\end{array}
\end{equation}
The above cancellation occurs actually for any number of generations.
However, it should be emphasized that the antisymmetric structure
itself in Eqn.(16) and others does not guarantee the above cancellation.
The crucial
point is that the dependence on quark flavours $d_k$ and $d_j$ is
completely separate. We believe that this point is also responsible for
the strong cancellation witnessed in the three loop QCD corrections
in the SM.
We also notice in passing that this cancellation is weaker than in
the SM case where it occurs even before summation over flavours if
one works in the unitarity gauge: Fig. 1(a) vanishes automatically
and others cancel among themselves due to simple equalities
$\cite{ck}$. In the present case however, Fig. 1(a) always
contributes, e. g. as indicated in the subsection 2B, and there are
no similar equalities which would demand the cancellation before
summation over flavours is taken.
Below we examine how this separate structure could be possible
for general couplings $C_{\alpha i}$ and $C_{\alpha i}^{\prime}$.
In other words, we want to determine what kinds of couplings are allowed
for the separate structure to occur.

Let us begin with the case of double $H^{\pm}$ exchanges. It is natural
to require from Eqn.(16) that
$C_1(e\alpha;kj),~C_2(e\alpha;kj),~C_4(e\alpha;kj)$ and
$C_5(e\alpha;kj)$ be independent of $d_k$ and $d_j$ and that
$C_3(e\alpha;kj)$ can only depend on $d_k$ and $C_3(e\alpha;jk)$
only on $d_j$. It is reasonable to assume that $C_{\alpha k}$ and
$C_{\alpha k}^{\prime}$ are universal as functions of quark masses
$d_k$ and $u_{\alpha}$, and that these masses do not appear as
denominators in $C_{\alpha k}$ and $C_{\alpha k}^{\prime}$. Then,
we must have $C_1(e\alpha;kj)=0$, so that
\begin{equation}
\begin{array}{l}
\displaystyle
(C_{ek}C_{ej}-C_{ek}^{\prime}C_{ej}^{\prime})
(C_{\alpha k}^{\prime}C_{\alpha j}-C_{\alpha k}C_{\alpha j}^{\prime})=
(C_{\alpha k}C_{\alpha j}-C_{\alpha k}^{\prime}C_{\alpha j}^{\prime})
(C_{ek}^{\prime}C_{ej}-C_{ek}C_{ej}^{\prime}),\\
\displaystyle C_2(e\alpha;kj)=2
(C_{ek}C_{ej}-C_{ek}^{\prime}C_{ej}^{\prime})
(C_{\alpha k}^{\prime}C_{\alpha j}-C_{\alpha k}C_{\alpha j}
^{\prime})\frac{1}{d_k^2-d_j^2},\\
\displaystyle C_3(e\alpha;kj)=-d_k^2C_2(e\alpha;kj),\\
\displaystyle
C_3(e\alpha;jk)=C_3(e\alpha;kj)|_{j\leftrightarrow k}.
\end{array}
\end{equation}
The independence in $C_2$ of $d_k$ and $d_j$ along with assumptions
about $C_{\alpha j}$ and $C_{\alpha j}^{\prime}$ implies,
\begin{equation}
\displaystyle
C_{\alpha k}C_{\alpha j}-C_{\alpha k}^{\prime}C_{\alpha j}^{\prime}
=\eta_{\alpha}(d_k+d_j),~~
C_{\alpha k}^{\prime}C_{\alpha j}-C_{\alpha k}C_{\alpha j}^{\prime}
=\delta_{\alpha}(d_k-d_j),
\end{equation}
where $\eta_{\alpha}$ and $\delta_{\alpha}$ are independent of $d_k$
and $d_j$ and may depend on $u_{\alpha}$. Furthermore, from the above
equations, we have
\begin{equation}
\displaystyle
(C_{\alpha k}+C_{\alpha k}^{\prime})(C_{\alpha j}-C_{\alpha j}
^{\prime})
=\eta_{\alpha}(d_k+d_j)+\delta_{\alpha}(d_k-d_j).
\end{equation}
The factorized dependence on $d_k$ and $d_j$ on the left-hand side
means that we may have two choices,
\begin{equation}
\begin{array}{ll}
\displaystyle
{\rm Case~(I):}&\eta_{\alpha}=+\delta_{\alpha},\\
\displaystyle &
C_{\alpha k}+C_{\alpha k}^{\prime}\propto d_k,~~
C_{\alpha k}-C_{\alpha k}^{\prime}~{\rm independent~of}~d_k;\\
\displaystyle
{\rm Case~(II):}&\eta_{\alpha}=-\delta_{\alpha},\\
\displaystyle &
C_{\alpha k}-C_{\alpha k}^{\prime}\propto d_k,~~
C_{\alpha k}+C_{\alpha k}^{\prime}~{\rm independent~of}~d_k
\end{array}
\end{equation}
Then,
\begin{equation}
\begin{array}{rcl}
\displaystyle C_4(e\alpha;kj)&=&\delta_e
(C_{\alpha k}\mp C_{\alpha k}^{\prime})
(C_{\alpha j}\mp C_{\alpha j}^{\prime}),\\
\displaystyle C_5(e\alpha;kj)&=&\delta_{\alpha}
(C_{ek}\mp C_{ek}^{\prime})
(C_{ej}\mp C_{ej}^{\prime}),\\
\end{array}
\end{equation}
where the upper sign corresponds to the Case (I) and the lower sign to
the Case (II). Therefore $C_4(e\alpha;kj)$ and $C_5(e\alpha;kj)$ are
both independent of $d_k$ and $d_j$. As one may have realized, MSSM
falls into the Case (I). Generally, the Case (I) corresponds to a
left-handed theory in the sense that the Higgs bosons are doublets
under $SU(2)_L$, while the Case (II) corresponds to a right-handed
theory under $SU(2)_R$. As far as the contribution from double
$H^{\pm}$ exchanges is concerned, the two cases are equivalent
up to a sign. This two-fold ambiguity is dismissed
when contributions from mixed exchanges of $H^{\pm}$ and $W^{\pm}$
are considered, since $W^{\pm}$
couples only to the left-handed current. Consider for example the
case of the subsection 2C. The coupling combination in Eqn.(23) is
$(\eta_{\alpha}+\delta_{\alpha})$ while those in Eqn.(25) become
\begin{equation}
\begin{array}{rcl}
\displaystyle C_1(e\alpha;kj)&=&2u_ed_kd_j
(\eta_{\alpha}-\delta_{\alpha}),\\
\displaystyle C_2(e\alpha;kj)&=&-2u_e
(\eta_{\alpha}+\delta_{\alpha}),\\
\displaystyle C_3(e\alpha;kj)&=&2u_ed_k
[\eta_{\alpha}(d_k+d_j)+\delta_{\alpha}(d_k-d_j)],\\
\displaystyle C_4(e\alpha;kj)&=&-2u_e
(C_{\alpha k}-C_{\alpha k}^{\prime})
(C_{\alpha j}-C_{\alpha j}^{\prime}),\\
\displaystyle C_5(e\alpha;kj)&=&2
[\eta_{\alpha}(u_e^2-d_kd_j)+\delta_{\alpha}(u_e^2+d_kd_j)].
\end{array}
\end{equation}
The previous requirements on these combinations then single out the
Case (I) as the only choice, otherwise the contribution in the
subsection 2C cannot be cancelled upon summing over quark flavours
and the quark EDM already arises at two loop order from mixed
exchanges of $H^{\pm}$ and $W^{\pm}$. The subsection 2D does not give
further constraints.

The Case (I) may be parametrized as follows,
\begin{equation}
\displaystyle C_{\alpha k}=x_{\alpha}d_k+y_{\alpha},~~
C_{\alpha k}^{\prime}=x_{\alpha}d_k-y_{\alpha},
\end{equation}
where $x_{\alpha}$ and $y_{\alpha}$ are independent of $d_k$ but may
depend on $u_{\alpha}$. A similar analysis may be repeated for the EDM
of the down-type quark $d_e$, following the prescriptions in Eqn.(18).
The cancellation of contributions from both double $H^{\pm}$ exchanges
and mixed $H^{\pm}-W^{\pm}$ exchanges requires,
\begin{equation}
\displaystyle C_{\alpha i}-C_{\alpha i}^{\prime}\propto u_{\alpha},~~
C_{\alpha i}+C_{\alpha i}^{\prime}{\rm ~independent~of~} u_{\alpha}.
\end{equation}
The constraints from Eqns.(39) and (40) then demand,
\begin{equation}
\displaystyle C_{\alpha i}=xd_i+yu_{\alpha},~~
C_{\alpha i}^{\prime}=xd_i-yu_{\alpha},
\end{equation}
where $x$ and $y$ are real constants that depend on the detail of the
models and cannot be determined from this analysis. We see that in
order to have a separate structure in the contribution to the quark
EDM, the $H^{\pm}$ couplings with quarks cannot be arbitrary, but have
to be in a form that is required by spontaneous symmetry breaking and
the Yukawa couplings.

\section{Conclusions}

We have presented an explicit two loop calculation of the quark EDM in
the two Higgs doublet extension of the SM where CP noninvariance
is assumed to be encoded in the nonzero phase of the CKM matrix.
Naively we would expect a contribution which is much larger than
the SM result arising at three loop order. However,
this large contribution is not realized in practice.
Our detailed calculation shows that the contribution involving
exchange of charged Higgs bosons vanishes completely due to
a cancellation mechanism.

We found that two factors are responsible for this complete
cancellation. One is the unitarity of the CKM matrix, the other is
the separate dependence in the relevant amplitude on the masses of
internal quarks which are weak doublet partners of the external quark
considered. We noticed that the antisymmetric structure itself of the
amplitude in internal quark flavours is sufficient to guarantee the
reality of the EDM, but is not to guarantee its complete cancellation.
To analyse the cancellation mechanism in more detail, we examined the
inverse problem of what kinds of couplings between $H^{\pm}$ and quarks
are allowed for the separate structure to occur. We found that the
couplings must be in a form that is dictated by spontaneous
symmetry breaking and the Yukawa coupling. This is consistent with
our assumption that the origin of the CP noninvariance resides in the
complex CKM matrix.

>From the vanishing result of the quark EDM at the lowest nontrivial
order, it is safe to conclude that it should be difficult to
detect the CP noninvariance through the EDM of the neutron or the
leptons if the CP noninvariance is of the CKM origin.

\noindent{\large Acknowlegements}

We thank Y.-L. Wu for conversations.

\appendix

\section{Some functions appearing in one loop diagrams}

The functions $f_i~(i=0,~1)$ appearing in renormalization of one loop
elements are
\begin{equation}
\begin{array}{rcl}
f_i(\ell^2,m^2,M^2)&=&\displaystyle \Gamma(\epsilon)(4\pi\mu^2)^{\epsilon}
\int_0^1dx~x^i[h(\ell^2,m^2,M^2;x)]^{-\epsilon},\\
h(\ell^2,m^2,M^2;x)&=&\displaystyle xM^2+(1-x)m^2-x(1-x)\ell^2,
\end{array}
\end{equation}
where $m$ and $M$ are respectively the masses of the quark and the
boson in the loop. The on-shell subtraction then produces the functions
$I_0$ and $I_1$ which depend on $d_k^2$ or $d_j^2$, as well as on
$\ell^2,~m^2,M^2$. These functions generally involve the function $h$
to the power of $-\epsilon$; but in the SEMA, it is sufficient to
expand them to the zero order in $\epsilon~(i=0,~1)$:
\begin{equation}
\displaystyle I_i(k)=I_i(\ell^2,m^2,M^2;d_k^2)=\int_0^1dx~x^i
\ln\frac{h(\ell^2,m^2,M^2;x)}{|h(d_k^2,m^2,M^2;x)|}+O(\epsilon).
\end{equation}
The functions $J_{i,j}(i=0,~1,~2;~j=0,~1)$ arise from differentiation with
respect to $\ell^2,~m^2$ or $M^2$ of the functions $f_i$:
\begin{equation}
\displaystyle J_{i,j}=J_{i,j}(\ell^2,m^2,M^2)=\int_0^1dx~
\frac{x^i(1-x)^j}{h(\ell^2,m^2,M^2;x)}+O(\epsilon).
\end{equation}

\begin{flushleft}
{\large Figure Captions }
\end{flushleft}

\noindent
Fig. 1  Diagrams that contribute to the EDM of the up-type quark
$u_e$. The wavy lines represent the electromagnetic fields and
the dashed lines the $H^{\pm}$ or $W^{\pm}$ fields. Diagrams for
the down-type quark $d_e$ are similar, with the replacements:
$\alpha\to i$ and $j,~k\to\alpha,~\beta$.

\newpage
\newpage
\begin{center}
\begin{picture}(400,600)(0,0)
\SetOffset(10,80)\SetWidth{2.}

\SetOffset(0,400)
\ArrowLine(20,100)(45,100)
\ArrowLine(45,100)(70,100)
\ArrowLine(70,100)(130,100)
\ArrowLine(130,100)(155,100)
\ArrowLine(155,100)(180,100)
\DashCArc(100,100)(55,180,360){4.}
\DashCArc(100,100)(30,0,180){4.}
\Photon(100,45)(100,20){3}{3}
\Text(33,111)[]{$e$}
\Text(58,111)[]{$j$}
\Text(100,111)[]{$\alpha$}
\Text(143,111)[]{$k$}
\Text(168,111)[]{$e$}
\Text(28,85)[]{$\displaystyle p+\frac{q}{2}$}
\Text(173,85)[]{$\displaystyle p-\frac{q}{2}$}
\Text(112,35)[]{$q$}
\Text(100,-5)[]{\large $(a)$}
\SetOffset(0,-400)

\SetOffset(200,400)
\Line(20,100)(180,100)
\DashCArc(100,100)(55,180,360){4.}
\DashCArc(100,100)(30,0,180){4.}
\Photon(100,130)(100,155){3}{3}
\Text(100,-5)[]{\large $(b)$}
\SetOffset(-200,-400)

\SetOffset(0,200)
\Line(20,100)(180,100)
\DashCArc(100,100)(55,0,180){4.}
\DashCArc(100,100)(30,0,180){4.}
\Photon(100,100)(100,75){3}{3}
\Text(100,45)[]{\large $(c)$}
\SetOffset(0,-200)

\SetOffset(200,200)
\Line(20,100)(180,100)
\DashCArc(100,100)(55,0,180){4.}
\DashCArc(110,100)(30,0,180){4.}
\Photon(63,100)(63,75){3}{3}
\Text(100,45)[]{\large $(d)$}
\SetOffset(-200,-200)

\SetOffset(0,0)
\Line(20,100)(180,100)
\DashCArc(100,100)(55,0,180){4.}
\DashCArc(90,100)(30,0,180){4.}
\Photon(137,100)(137,75){3}{3}
\Text(100,45)[]{\large $(e)$}
\SetOffset(0,0)

\SetOffset(200,10)
\Text(5,5)[]{\large Figure $(1)$}

\end{picture}\\
\end{center}

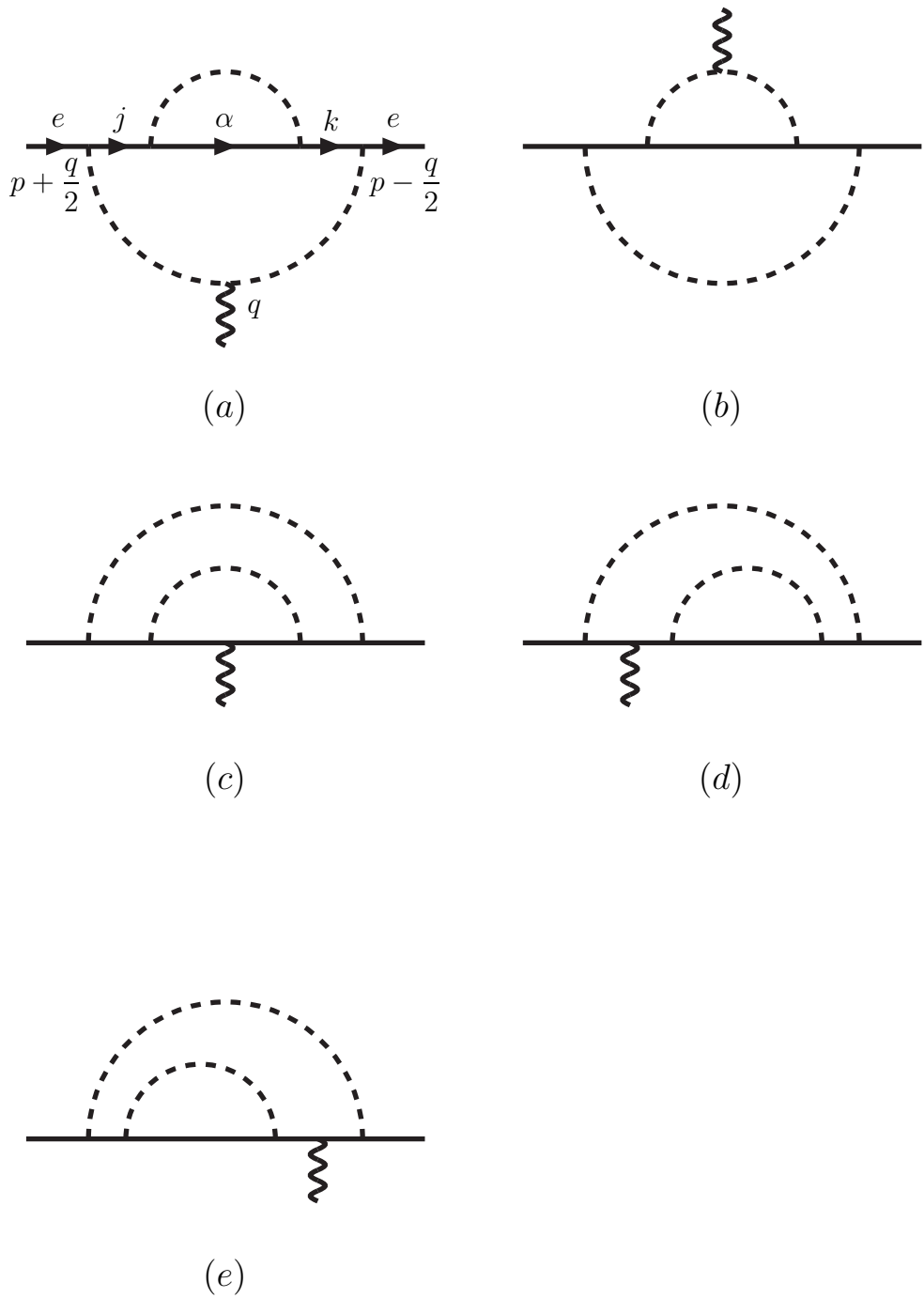
\end{document}